\documentclass{iopart}

\usepackage{epsf}

\newcommand{\insertplot}[3]{
	\begin{figure}
	  \begin{center}
 		\leavevmode\epsfysize=120mm \epsfbox{#1}
	  \end{center}
	  \caption{\label{#2} #3}
	\end{figure}
}

\newcommand{\be}{\begin{equation}}
\newcommand{\ee}{\end{equation}}
\newcommand{\ba}{\begin{eqnarray}}
\newcommand{\ea}{\end{eqnarray}}

\newcommand{\vs}{\vspace{5mm} }

\begin{document}


\title{
Quark coalescence in the mid rapidity region at RHIC
}

\author{ T. S. Bir\'o\footnote[2]{Talk given at SM'2001 Frankfurt, Germany},  
P. L\'evai and J. Zim\'anyi
}

\address{
 KFKI Research Institute for Particle and Nuclear Physics, \\ 
  P. O. Box 49, Budapest, H-1525, Hungary \\
}

\begin{abstract}

We utilize the ALCOR model for mid-rapidity hadron number
predictions at AGS, SPS and RHIC energies. We present simple 
fits for the energy dependence of stopping and quark production.

\end{abstract}



\section{Introduction}
\label{INTRO}

\vs
By comparing the assumptions and predictions of different models
for describing relativistic heavy ion collisions, each among others and
with experiments, it is time to make this comparison throughout
a wide energy range, from AGS via SPS to RHIC and LHC energies.
This is crucial both for pinning down the question of
transition between qualitatively different behaviors, 
dominated by hadron -- hadron or parton -- parton
reactions, respectively.
Investigating the energy dependence is as well important
to find differences in the agreement of different models with
the trends revealed by a vast amount of new experimental data.

\vs
One of the longest standing debate of the last years is that
between followers of the equilibrium concept and pursuer of
micro-dynamical approaches. 
Considering quark coalescence our ALCOR model
\cite{Bir1,ALCOR1, ALCOR2} belongs to the second party. Even
if one would allow for less detailed models by making predictions,
the global chemical equilibrium hypothesis 
\cite{Braun-Munziger,BM2,Cleymans}
in our oppinion is ruled out. It is in particular worth to
mention here, that the solely fact, that certain particle
ratios are rapidity dependent, contradicts to global chemical equilibrium.
In fact a RHIC experiment has found different
anti-proton to proton ratios at mid rpidity and moderate rapidity:
$\overline{p}/p = 0.64$ at $y=0$ and $\overline{p}/p = 0.41$ at
$y=2$ \cite{BRAHMS}.

\vs
Taking this experimental fact into account we utilize now
the formerly global quark matter hadronization model
ALCOR to the mid-rapidity window only.
The underlying assumption is that there is no significant
exchange of flavor between far lying rapidity intervals
(this is a reminder to the Bjorken flow picture and to
different versions of string and color rope models).
The unit width mid-rapidity window is selected for comparison
of the predictions of the modified ALCOR model with 
experimental data at AGS, SPS and RHIC energies.
The AGS data are presented rather for estimating the tendency
of an assumption of constituent quark matter formation
underlying the ALCOR model down to low energies, we actually
expect this model to fail at AGS.

\vs
In this talk we shall concentrate on the energy dependence
of the ALCOR model parameters extracted from fits to a certain
few experimental observations (and checking against several
others). In order to appreciate these dependencies we
briefly discuss the physical picture behind ALCOR and the
way how the model parameters are fixed.
Finally we sketch some speculations about possible
causes for the obtained energy dependence.

\section{Out of equilibirum}
\label{NON-EQ}

\vs
The name of the ALCOR model stems from the abbreviation of
an $A$lgebraic $Co$alescence $R$ehadronization model.
Hadron formation by coalescence assumes a constituent quark matter
before hadronization and a fast process relative to 
typical interaction times after the hadron formation.
In fact ALCOR neglects secondary collisions after hadronization
but includes hadronic resonance decays. In the presented version
the quark clusters belong to the ground state and to the
first excited state hadrons with equal probability
(naturally counting spin degeneracies) and all decay to
the ground state mesons and barions before detection.

\vs
Due to its basic assumption ALCOR is a {\em statistical}
model: it considers average events and manipulates
with average total numbers in a given section of the phase space
(mid-rapidity unit window presently). It is a {\em coalescence}
model: it describes mesonic prehadrons as $q\overline{q}$
quark -- antiquark clusters with varying flavor composition
and baryonic ones composed from three quarks ($qqq$),
antibaryonic clusters correspondingly from three antiquarks.
ALCOR concentrates to the {\em chemical} evolution:
no chemical potentials are used and no chemical equilibrium
is assumed at any instance of the calculation.
Finally ALCOR utilizes a picture of thermal but largely massive
{\em quark matter} without gluons at the beginning of the
fast hadronization.

\vs
Perhaps it is as interesting to list what ALCOR is not.
It is not an {\em equilibrium} model, because it relies
on branching ratios calculated from in-medium cross-sections.
It is not a {\em linear} coalescence model, because
it uses extra factors, the $b_i$-s, by quark counting
in order to take into account concurrence between
different hadronization channels. It is also not a {\em transport}
model, because it assumes a thermalized state to begin with
at a given temperature, $T$, the spectra of the finally observed hadrons
reflect this thermalized massive constituent quark matter
(after an additional free streaming) by assumption.
This is one of the major simplifications of the model,
this assumption can be tested by comparing to transport models.
Finally ALCOR does not assume {\em quark-gluon plasma}
as a state before hadronization, it actually leaves open the
question from what the constituent quark matter is formed from
and how has it been thermalized.

\section{Quark matter or hadron matter}
\label{QM-HM}

\vs
We would like to stress here again that the concept of
``quark matter'' we are using in ALCOR is not that of
the original free plasma of non-interacting, massless quarks
and gluons. In line with the experience gained by studying
the interacting QCD plasma with different methods by a great
number of people, in particular considering the phenomena of
static screening, of massive dispersion relations and
the observation hadron-like correlated few-quark-clusters
in lattice QCD simulations, we utilize the following
picture of ``prehadron matter'': it contains very few gluons,
quarks and antiquarks occur and interact with large effective
masses near to the constituent mass value, and a strong,
string-like interaction -- also shown in the magnetic
area-law in some lattice calculations -- bounds these
quarks into color neutral clusters with hadronic quantum numbers.

\vs
According to this picture the main assumptions of the ALCOR
model can be summerized as follows:
\begin{itemize}
\item	new quark -- antiquark pairs are produced before
	the hadronization process,
\item	all gluons and gluonic fields fragment into quark -
	antiquark pairs,
\item	and therefore the final hadronization is practically
	a {\em redistribution} of quarks and antiquarks
	among the possible clusters,
\item	the final hadrons have the same flavor composition
	as the prehadronic quark clusters.
\end{itemize}

\section{ALCOR predictions and energy dependence}
\label{RESULTS}

The input parameters of the ALCOR model can be divided to two 
categories:
in the first belong those which are restricted by rational
estimates and knowledge about elementary processes.
The constituent quark masses we use are designed to describe
static hadron properties (we use $m_u=m_d=300$ MeV, $m_s=500$ MeV).
The branching ratios we calculate on the basis of assumed
quark matter properties, we use a strong coupling $\alpha_S=0.85$,
a wave packet size of $\rho=0.3$ fm and a medium temperature
of $T = 170$ MeV.


\insertplot{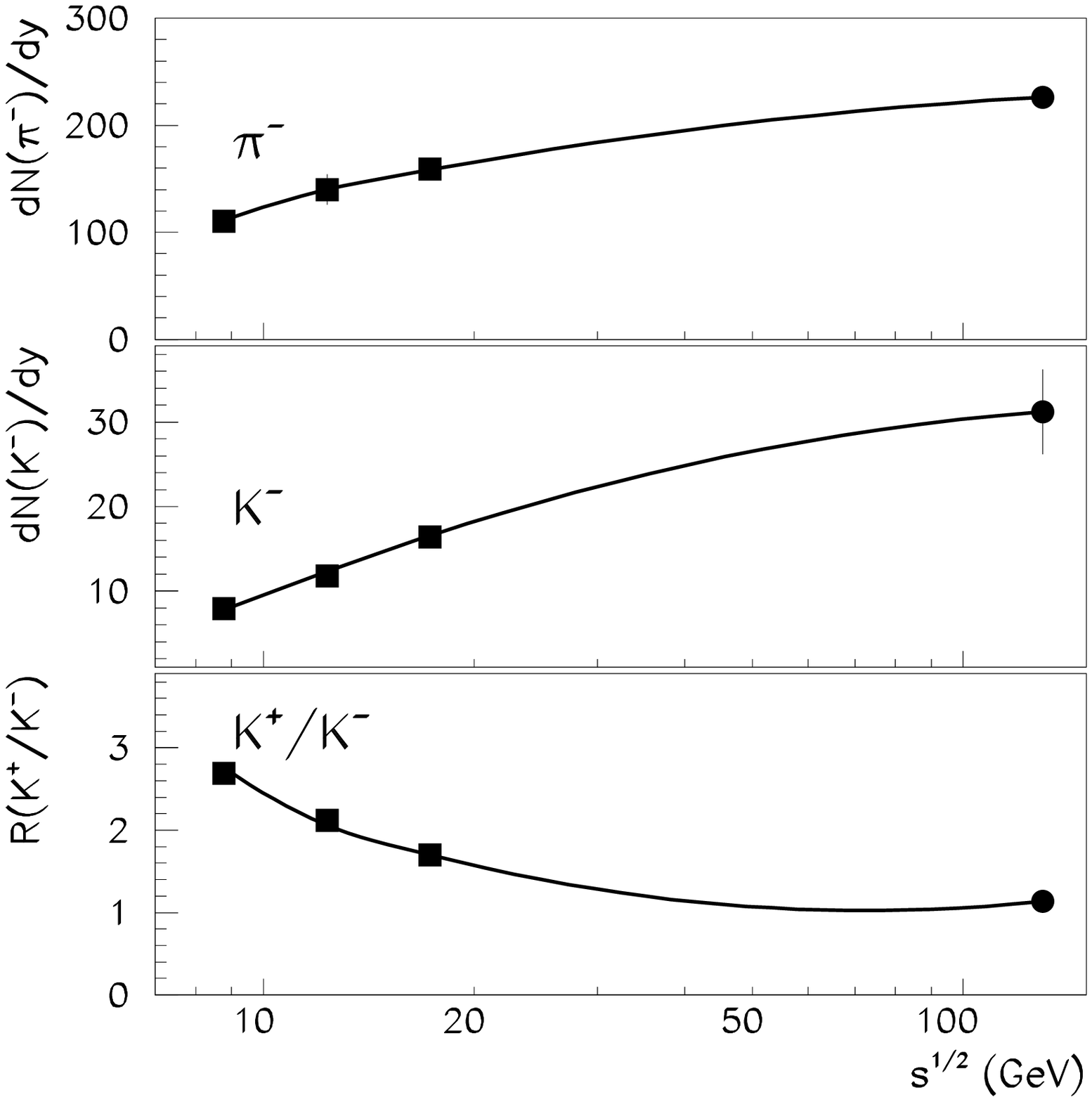}{EXP-DATA}{The energy dependence
of selected experimental heavy-ion data used for ALCOR input.}

\vs
In the second category belong those parameters which are obtained from
fitting the results of ALCOR to experimental heavy-ion data.
We have three such parameters: the stopping, here defined
by the baryon per cent in the mid-rapidity window stemming from
the colliding nuclei, is fitted to the experimental $K^+/K^-$
ratio, and two characteristic parameters of meson production,
namely the number of produced light and strange quark pairs,
$N_{u\overline{u}}$ and 
$f_s=N_{s\overline{s}}/(N_{u\overline{u}}+N_{d\overline{d}})$,
are fitted to the $\pi^-$ and $K^-$ yield, respectively.
These experimental data are displayed in Fig.\ref{EXP-DATA}
as a function of the bombarding energy.

\vs
The results incorporate total and mid-rapidity numbers of
elements of
the lowest lying meson, baryon and antibaryon multiplets
in the SU(3) flavor space, spectra with an assumed flow
pattern (in the MICOR version \cite{MICOR}), 
and predictions to LHC energy using
the fitted energy dependence of the parameters.


\insertplot{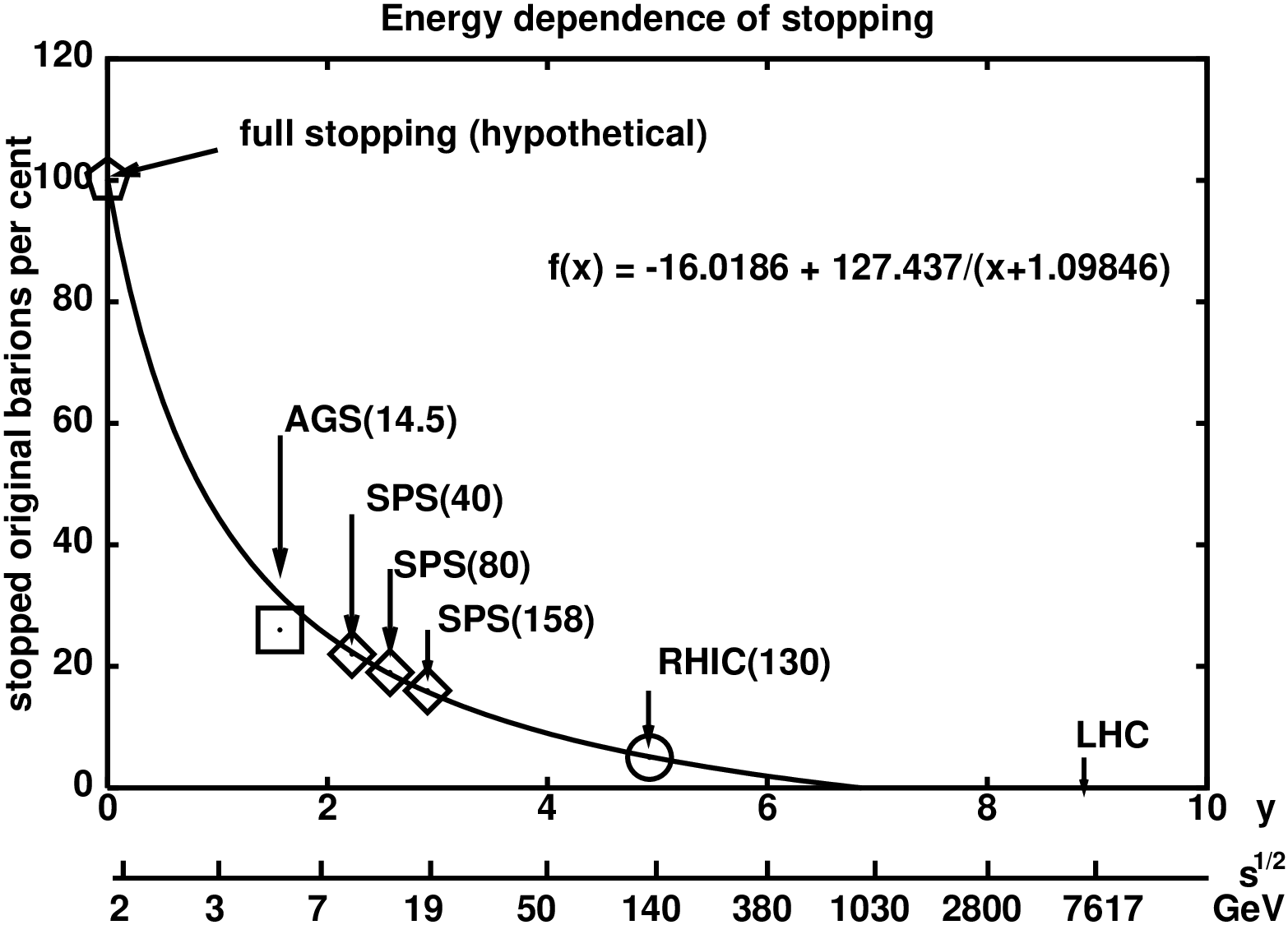}{STOP}{The energy dependence of the stopping
power as obtained from ALCOR by fitting the $K^+/K^-$ yield.}

Let us start with the energy dependenc of the stopping (cf. Fig.\ref{STOP}).
Here a decrasing tendency of the stopping power effective to
the mid-rapidity window is insepcted alone from the SPS and RHIC
data. Assuming full stopping at low energy, which is a hypothetical
point, the best fit reaches zero stopping already at a finite
energy, before the LHC regime. The number of stopped baryons
in the observed range roughly scale with the energy as the elastic
cross section vanishes.
The AGS point is somewhat below the fitted curve, but we do not
expect ALCOR to be realistic at such a low energy.

\insertplot{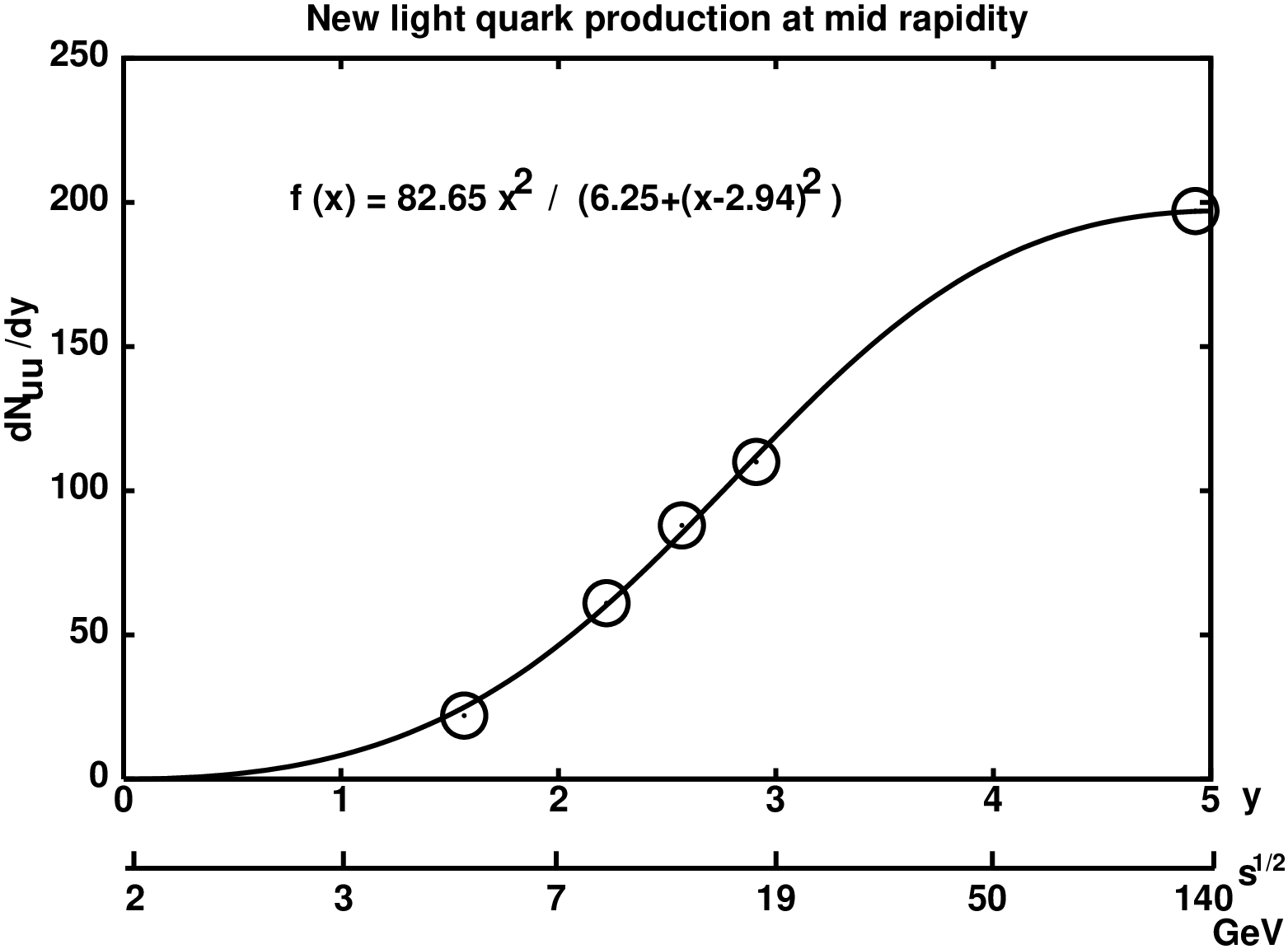}{QQ}{The energy dependence of the newly produced
quark pairs as obtained from ALCOR by fitting the $\pi$ yield.}

Inspectnig Fig.\ref{QQ} it is easy to realize that
the energy dependence of the newly produced quark -- antiquark
pairs seems to saturate around RHIC energy. (This is, however,
only one possible fit to the present data, in this respect
the LHC result will be crucial.) This behavior agrees with
Regge double-pole estimates for the $pp$ cross section at very high energies:
$N_{u\overline{u}} \sim \sigma^{tot} \sim A \, {{\rm ln}\; s} + B$
\cite{ESTIMATE}.
Since at high energy the relative rapidity is 
$y \approx {{\rm ln} \; s },$
the number of new pairs in the mid-rapidity window,
$dN/dy\approx N_{u\overline{u}}/y \approx A + B/({\rm ln} \; s)$ saturates.

\insertplot{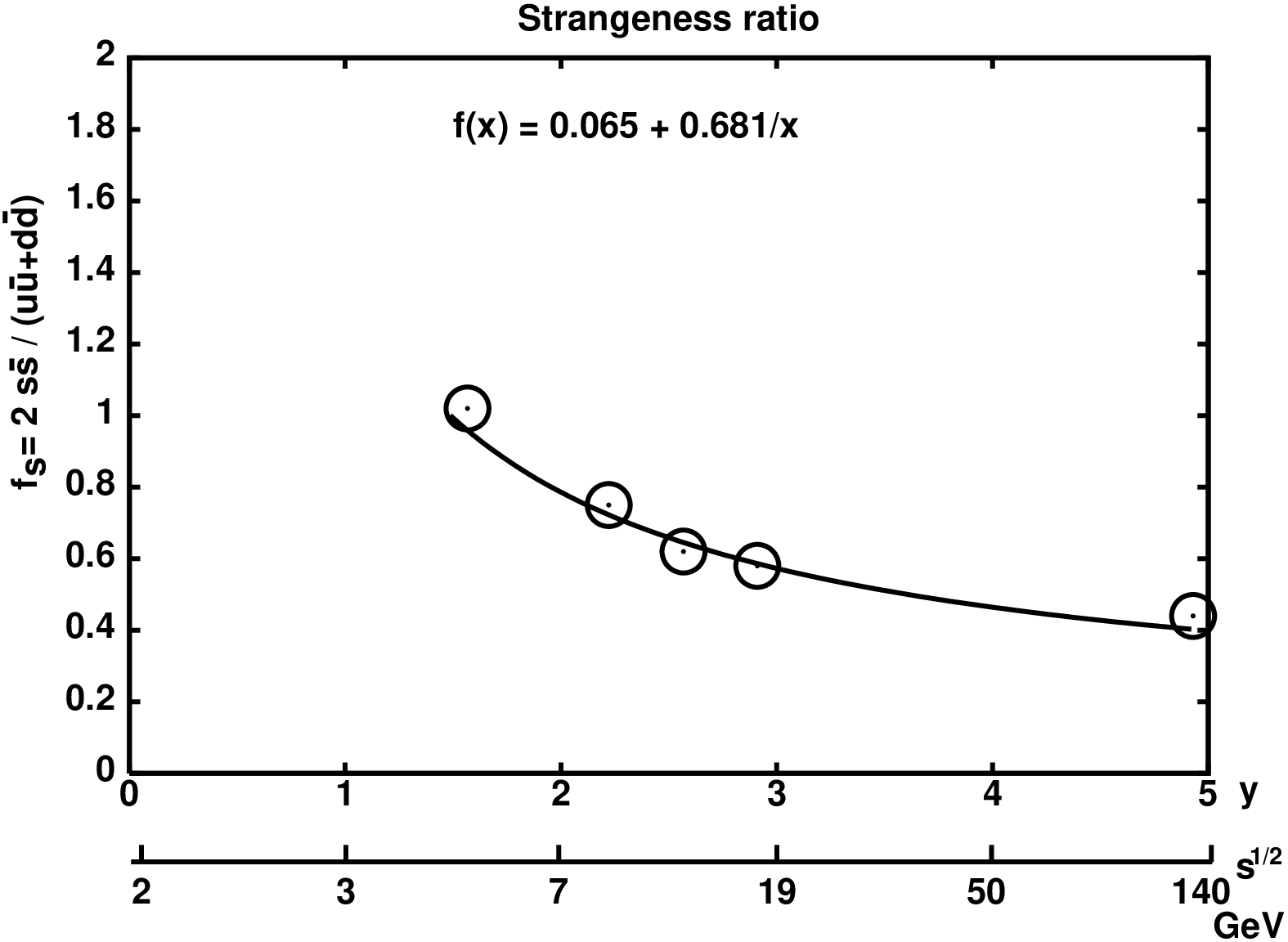}{FS}{The energy dependence of the starngeness
ratio as obtained from ALCOR by fitting the $K/\pi$ yield.}

The strangeness ratio, shown in Fig.\ref{FS} 
also falls with increasing bombarding energy.
The saturation value at high energy agrees with estimates
based on elementary $pp$ physics, the rise towards lower
energies requires explanation. In our oppinion it has to do
with the fact that the strange constituent mass is bigger than
the up and down masses, and therefore strange hadrons show a
reduced rapidity dispersion at low bombarding energies.
This may occur as an increasing ratio at mid-rapidity.

\insertplot{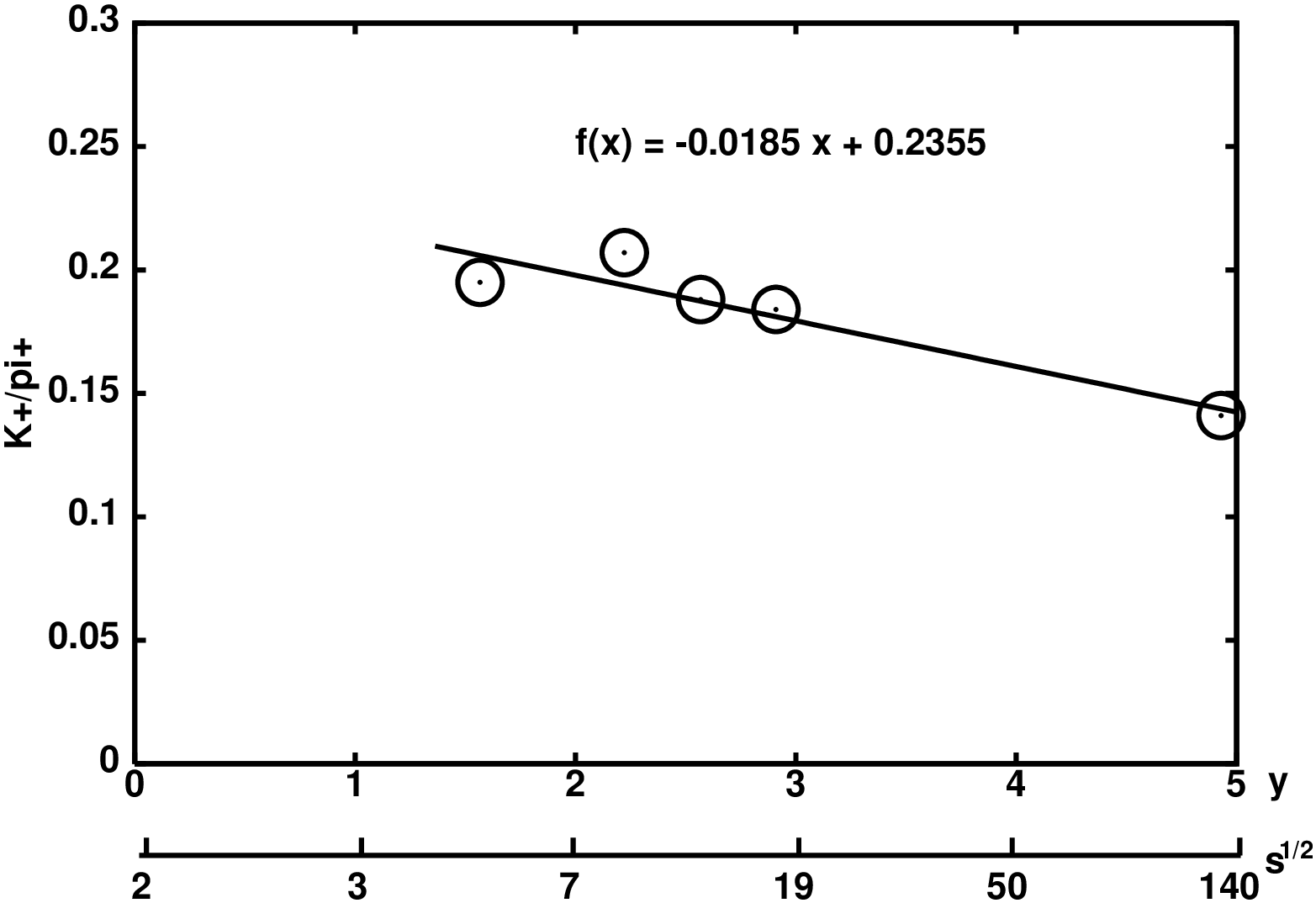}{KPI}{The energy dependence of the $K^+/\pi^+$
ratio fitted by ALCOR.}

Finally, as an example, the energy dependence of the
$K^+/\pi^+$ ratio is shown in Fig.\ref{KPI}. 
The encircled points are ALCOR
results at each respective energy, the line shows a fitted,
linear function to these points. Reflecting the decrease
of the strangeness ratio in the prehadron matter, this hadronic
ratio is also decreasing with increasing bombarding energy.

\begin{table}
\begin{center}
\begin{tabular}{llll}
\hline
&ALCOR model & Preliminary data & Ref. \\ \hline\hline
 $h^- $ 
& 260        & $264 \pm 18$           & STAR \cite{STAR4}   \\
 ${K^+}/{\pi^+}$
& 0.142      & $0.15 \pm 0.01$              & STAR \cite{STAR13}   \\
 ${K^+}/{K^-}$
& 1.13       & $1.12 \pm 0.06$              & STAR \cite{STAR4}   \\
 ${\Xi^-}/{\pi^-}$
& 0.015      & $0.014 \pm 0.01$             & STAR \cite{STAR5}   \\
 ${\overline p}^-/{p}^+$
& 0.63       & $0.61 \pm 0.06$              & STAR \cite{STAR4}   \\
 ${\overline \Lambda}/{\Lambda}$
& 0.72       & $0.73 \pm 0.03$              & STAR \cite{STAR4}   \\
 ${\overline \Xi}^+/{\Xi}^-$
& 0.83       & $0.82  \pm 0.08$             & STAR \cite{STAR4}   \\
 $K^{*+}/h^-$
& 0.077      & 0.065                       & STAR \cite{STAR4}   \\
 ${\overline K}^{*-}/h^-$
& 0.067     & 0.060                        & STAR \cite{STAR4}   \\ \hline
\end{tabular}
\end{center}
\caption{
Hadron production in Au+Au collision at $\sqrt{s}=130$ AGeV
from the ALCOR model and the preliminary experimental data
\cite{STAR13,STAR4,STAR5}}

\end{table}

In conclusion
the measured data are in good agreement with ALCOR predictions.
We also transformed some of the measured particle numbers into physical
        quantities with the help of ALCOR, in order to gain
a better insight into the energy dependence of the stopping,
quark pair creation and strangeness ratio.
At RHIC we realize that 
we are near to the complete transparency which will
        be reached before LHC energy according to ALCOR.
At RHIC we are also near to the maximal prehadronization
        energy density (see number of new
        midrapidity ${q,{\overline q}}$ pairs in Table 1.).


\vs

\ack
This work was supported by the Hungarian Science Fund (OTKA) grants
T 34269.

\end{document}